\documentclass[prl,floatfix,twocolumn,superscriptaddress,aps,psfig]{revtex4} 

\newcommand{\Litwo}{Li$_2$CuO$_2$}
\newcommand{\Catwo}{Ca$_2$Y$_2$Cu$_5$O$_{10}$}
\usepackage{graphicx,color,epsfig,rotate,amssymb}

\begin{document}
%\draft

\title{Saturation field of frustrated chain cuprates: broad regions of  predominant interchain coupling}

\author{ S.~Nishimoto}

\author{S.-L.~Drechsler$^{*}$}

\author{R.O.\ Kuzian} 

\author{J.\ van den Brink}

\affiliation{IFW Dresden, P.O.~Box 270116, D-01171 Dresden, Germany} 

\author{J.~Richter}

\affiliation{Universit\"at Magdeburg, Institut f\"ur Theoretische Physik, Germany}

\author{W.E.A.~Lorenz}

\affiliation{IFW Dresden, P.O.~Box 270116, D-01171 Dresden, Germany} 

\author{Y.\ Skourski}

\affiliation{Hochfeld-Magnetlabor Dresden, Helmholtz-Zentrum Dresden-Rossendorf, D-01314 Dresden, Germany}

\author{R.~Klingeler}

\affiliation{Kirchhoff Institute for Physics, University of Heidelberg, D-69120 Heidelberg, Germany}

\author{B.~B\"uchner}

\affiliation{IFW Dresden, P.O.~Box 270116, D-01171 Dresden, Germany}

\date{\today}

\begin{abstract}

An efficient and precise thermodynamic method to extract the interchain coupling (IC) of 
spatially anisotropic 2D or 3D spin-1/2 systems from their empirical saturation field $H_{\rm s}$ ($T=0$) 
is proposed. Using density-matrix renormalization group, hard-core boson, and spin-wave theory
we study how $H_{\rm s}$ is affected by an antiferromagnetic (AFM) IC between frustrated chains 
described in the $J_1$-$J_2$-spin model with ferromagnetic 1st and 
%antiferromagnetic 
AFM
2nd neighbor 
inchain exchange. A complex 3D-phase diagram has been found. For 
%the reference systems 
\Litwo \ 
and \Catwo,we  show that 
%their 
$H_{\rm s}$ is solely determined by the IC 
and predict  $H_{\rm s} \approx 61$~T for the latter.
Using $H_{\rm s} \approx 55$~T 
from 
%reported here 
our 
high-field pulsed measurements one reads out a weak IC for \Litwo \ close 
to that from neutron scattering. 
%For \Catwo \ we predict $H_{\rm s} \approx 61$~T. 
\end{abstract}

%\pacs{74.25Bt, 74.25Op}

\maketitle

Since real spin chain systems exhibit besides a significant in-chain coupling also an 
{\it interchain} coupling (IC), one may ask: in which cases is this relatively weak IC 
still important or even crucial? From the Mermin-Wagner theorem its decisive role 
for the suppression of fluctuations is well-known. The IC leads to magnetic long-range 
order (LRO) at $T=0$ in 2D \cite{Sandvik99} and at $T < T_{\rm N}$ in 3D 
(see e.g.\ Ref.\ \onlinecite{Todo2008}).
Often one is faced with a situation that the (large) in-chain couplings are known with
reasonable precision, e.g.\ from band-structure calculations, inelastic neutron 
scattering (INS) or susceptibility data \cite{Lorenz09}, but precise values for 
the tiny (nevertheless important) IC are lacking.
If frustration is absent, the IC can be determined quite accurately, 
e.g.\ from $T_{\rm N}$ analyzed by Quantum Monte Carlo based studies \cite{Yasuda05} 
and more approximately using mean-field theory for a 3D IC \cite{Eggert02}. 
But how to extract from experimental data a small IC for frustrated systems 
with weakly coupled chains where these methods do not work? Here
we address such a 2D/3D problem for the case of frustrated spin-1/2 chains 
with ferromagnetic (FM) nearest-neighbor and antiferromagnetic (AFM) 
next-nearest-neighbor exchange described by the spin isotropic $J_1$-$J_2$-model 
(IM). Nowadays it is the standard  model for edge-shared chain cuprates (see e.g.\ \cite{Drechsler05}). 
This 1D-IM attracted special interest \cite{Chubukov91,Vekua07,Heidrich-Meisner06,Kuzian07,haertel08,Dmitriev09} 
due to a rich phase diagram with multipolar (MP) phases derived from multi-magnon bound states (MBS) 
in high magnetic fields \cite{Kecke07,Sudan09,Hikihara08}. Additional AFM degrees of freedom 
enhance the kinetic energy of magnons. Thus, AFM IC might disfavor multi-MBS. 
A precise 
knowledge of
%insight into 
the magnitude of the IC is therefore a necessary prerequisite 
to attack the multi-MBS problem, including a possible MBS Bose-Einstein condensation. 
Hence, knowing it is of general interest \cite{Zinke09,Ueda09,Zhitomirsky10,Svistov11}. 

\begin{figure}[b!]
\begin{center}
%\begin{minipage}{0.85\textwidth}
%\hspace{-7.5cm}
\includegraphics[width=0.41\textwidth,angle=0]{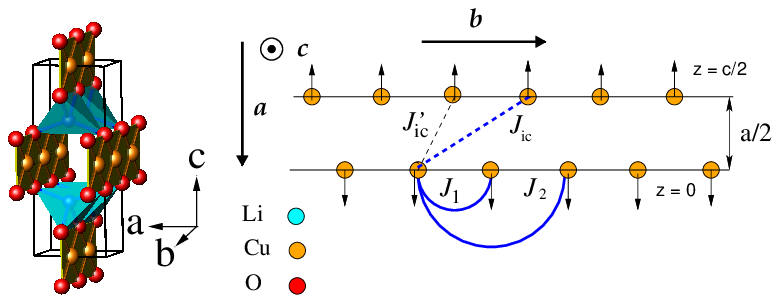}
%\end{minipage}
\end{center}
\caption{(Color) Left: Crystal structure of \Litwo \ with 
two 
%AFM coupled 
CuO$_2$ chains per unit cell along the {\it b}-axis. 
Right: View along the {\it c}-axis on the {\it ab}-plane.
The main in-
%chain 
and interchain couplings 
$J_{1,2}$ and $J_{\rm ic}$, $J^\prime_{\rm ic}$: 
%are marked by 
%blue 
arcs and 
dashed lines, respectively. 
The 
%We set 
normalized ICs 
read
%as 
$\beta_1=J^\prime_{\rm ic}/|J_1|$ and $\beta_2=J_{\rm ic}/|J_1|$. 
} 
\label{fig::Struct}
\end{figure}

In this context, \Litwo \ (see Fig.\ 1) is one of the best studied frustrated cuprates
and therefore well suited to compare theory and experiment. 
In particular, using spin-wave theory (SWT) the main in-chain and IC $J$'s were extracted 
from INS data and a specific AFM IC was found crucial for preventing spiral order 
in the 3D ground state (GS) \cite{Lorenz09}. If the saturation field $H_{\rm s}$ 
can be determined experimentally, an excellent opportunity occurs to check the INS 
based IC. But for \Litwo \ $H_{\rm s}$ has not been measured so far.
Here we report high-field magnetization data to fill this gap. Our paper is organized as follows.
First we revisit the 1D case and provide details of the density-matrix renormalization 
group (DMRG) technique involved. Then we report results for coupled chains, among them 
a rather complex phase diagram, and compare our findings with our experimental 
$H_{\rm s}$-data for \Litwo \ and arrive at an almost perfect explanation in terms of 
a predominant IC.\\

\begin{figure}[t]
\includegraphics[width=6.3cm]{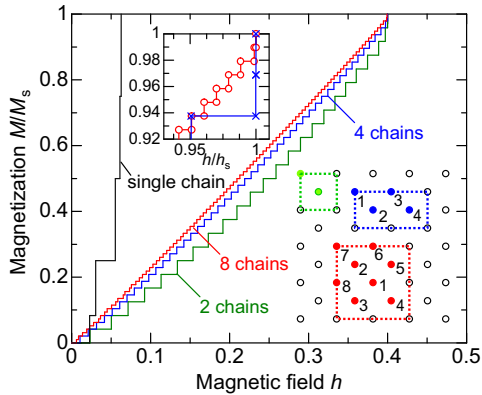}
\caption{(Color) Magnetization vs.\ external field (DMRG data) 
for different $n$ as shown in the inset, for $\alpha=J_2/|J_1|=1/3$, $\beta_2=0.05$, $\beta_1=0$, 
and $L=24$ sites in each chain. Lower inset: The 3D arrangement of chains used for 
the DMRG study reported here. Upper inset: Magnetization curve inside the 3-MBS region for 
$\beta_2=0.005$ (blue curve, \textcolor{blue}{$\times$}) compared with that for the 
1-magnon C-phase for $\beta_2=0.05$ (red curve, \textcolor{red}{$\circ$}). 
Note the three times larger step for $\beta_2=0.005$.
}
\label{magnetization}
\end{figure} 

\vspace{-0.05cm}

We apply the DMRG method~\cite{White92} with periodic boundary conditions (PBC) 
in all directions. 
%Usually it is believed that 
Seemingly, 
this method is much less favorable for $D>1$; 
however, on current workstations using highly efficient DMRG codes, spin systems 
with up to about $\sqrt{10} \times \sqrt{10} \times 50$ sites, i.e.\ 10 coupled 
chains of length $L \sim 50$, can be studied. Thus, by taking a proper arrangement 
of the chains 3D lattices can be simulated, cf.\ the inset in Fig.~\ref{magnetization}. 
Let us now describe how the block states are constructed: in the $n \times L$ cluster, 
where $n$ denotes the number of chains and $ L $ is the chain length. 
If we regard $n$ sites in the {\it ac}-plane as a ``unit cell'', the system can 
be treated as an effective 1D chain with $L$ sites (step 1). This enables us to use 
an appropriate 1D array for the construction of the PBC (see Fig.~1 of Ref.~\onlinecite{Qin95}). 
In the second step the sites within each ``unit  cell'' are arranged into numeric 
order as shown in the inset of Fig.~\ref{magnetization}. This way, the distance 
between most separated interacting sites can be held at $11$ and $23$ in 
the $4$- and $8$-chain systems, respectively. Since the exchange interactions 
run spatially throughout the system, the wave function 
converges very slowly with DMRG sweep but without getting trapped in a `false' 
GS. We typically kept $m \approx 1600-4000$ density-matrix eigenstates in the DMRG procedure. 
About $20-30$ sweeps are necessary to obtain the GS energy within a convergence of 
$10^{-7}|J_1|$ for each $m$ value. All calculated quantities were extrapolated 
to $m \to \infty$ and the maximum error in the GS energy is estimated 
as $\Delta E/|J_1| \sim 10^{-4}$, while the discarded weight in the renormalization 
is less than $1 \times 10^{-6}$. For high-spin states [$S_{\rm tot}^z \gtrsim (nL-10)/2)$] 
the GS energy can be obtained with an accuracy of $\Delta E/|J_1|< 10^{-12}$ 
by carrying out several thousands sweeps even with $m \approx 100-800$. 
Then, we obtain the reduced saturation field $h_{\rm s}=g \mu_{\mbox{\tiny B}} H_{\rm s}/|J_1|$ 
with high accuracy (e.g.\ 12 digits as compared to exact solutions available 
in some cases). The phase assignment in the 1D-diagrams shown in 
Refs.\ \onlinecite{epaps,Kecke07,Hikihara08,Sudan09,Dmitriev09} and in 
3D (see Fig.\ 3) stems from an analysis of the magnetization curve just 
slightly below $h_{\rm s}$ as shown in Figs.\ 2 and 6. The signature of 
each region is the height of the magnetization steps $\Delta S^z$=1,2,3, ... for 
dipolar (1-magnon), quadrupolar (2-MBS), octupolar (3-MBS), and 
hexadecupolar (4-MBS), etc.\ phases, respectively. 
For the quadrupolar case the DMRG results are confirmed by the exact hard-core boson 
approach (HCBA) \cite{Kuzian07}.

\begin{table}[t]
 \caption{Saturation field $h_{\rm s}$ at  $\alpha=0.332$, $\beta_2=3/76$, and 
$\beta_1=0$. $J_{\rm ic}$ has been multiplied by 4 for 2-chain systems.
}

\begin{center}
\begin{tabular}{ccccccc}
    \hline
    \hline
     $L$ &  single chain   & 2-chain system  & 8-chain system   &  \\
    \hline
   16  & 0.0610480058942 & 0.315789473684 & 0.315789473684 &  \\
   48  & 0.0616910423378 & 0.315789473684 & 0.315789473684 &  \\
   96  & 0.0616910487270 & 0.315789473684 &                &  \\
  144  & 0.0616910487247 & 0.315789473684 &                &  \\
    \hline
    \hline
  \end{tabular}
 \end{center}
\end{table}

\begin{figure}[b]
\includegraphics[width=5.8cm]{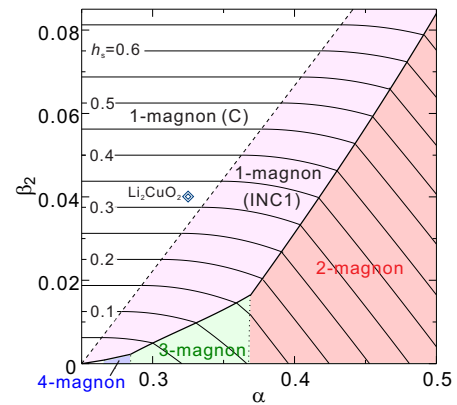}
\caption{(Color) Part of the 3D phase diagram 
around \Litwo \ in terms of 
%the IC
 $\beta_2$ 
and $\alpha$. $h_{\rm s}$ is given by contour lines.
}
\label{f3}
\end{figure}

\begin{figure}[b]
\includegraphics[width=6.4cm]{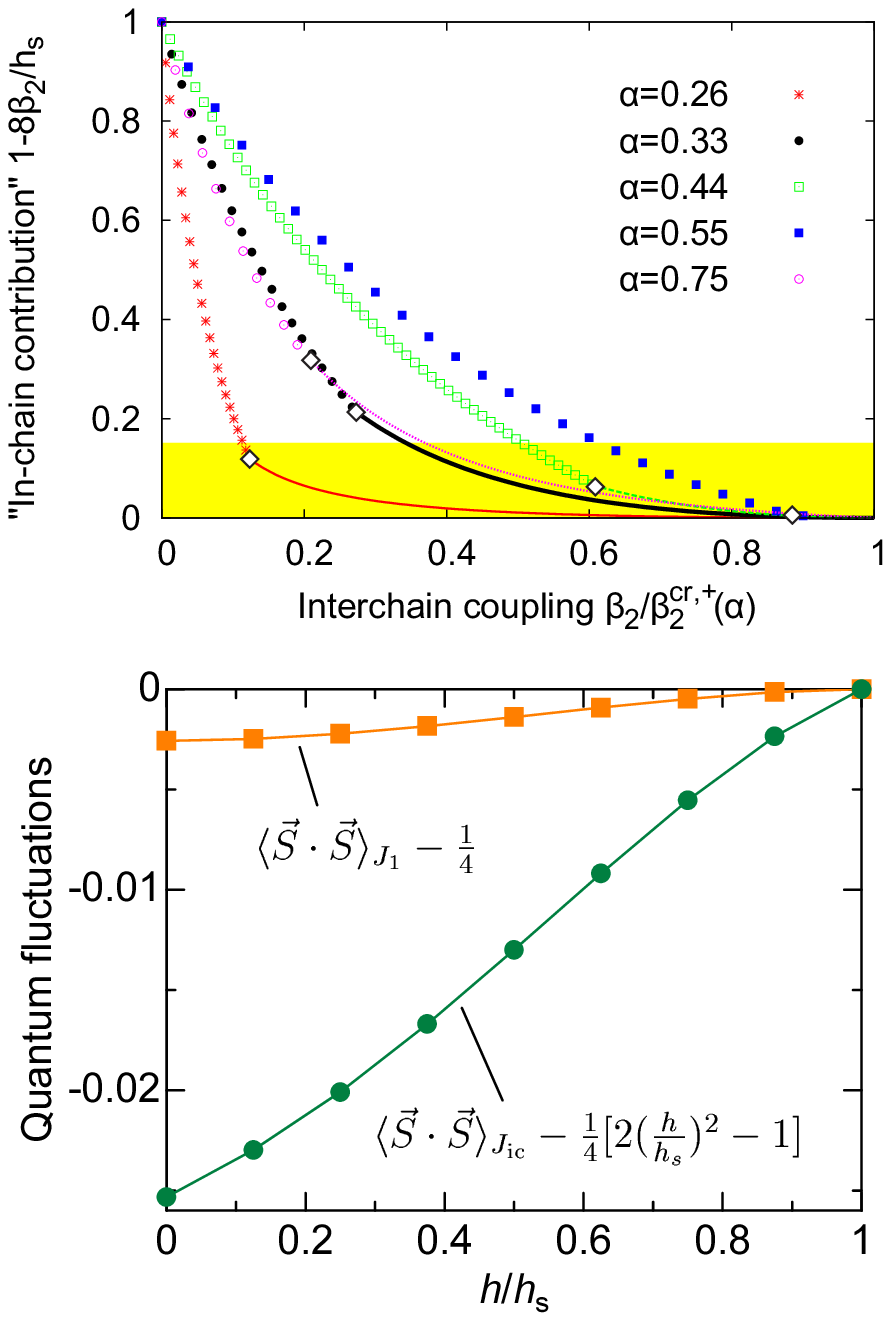}
\caption{(Color)
Upper: Approximate relative in-chain contribution to the saturation field 
measured by its deviation from Eq.\ (1) 
for different $\alpha$ values vs.\ $\beta_2$ in units of its critical 
value $\beta^{\rm cr,+}_2$ (see Figs.\ 3, S1, and text). The yellow stripe highlights the 
region of predominant IC. The full line parts of a curve (below the symbol $\diamond$) can 
be determined within the SWT, whereas the branches given by symbols are obtained by 
the DMRG and the HCBA methods. Lower: in-chain and IC spin-spin correlation functions 
vs.\ applied field, given by the filled symbols {\color[named]{Orange}$\blacksquare$} and 
{\Large \color[named]{Green}${ \bullet}$}, respectively, as obtained by DMRG.
}
\label{f4}
\end{figure}

We start with the 1D problem. Using its frustration parameter $\alpha$=$J_2/|J_1|$,  
the critical point is given by the level crossing of a singlet and the highest multiplet state 
at $\alpha_{\rm c}=1/4$. Approaching $\alpha_{\rm c}$ from the spiral side $\alpha \geq \alpha_{\rm c}$, 
it is clear that $H^{\rm 1D}_{\rm s}(\alpha)$ decreases and vanishes at $\alpha_{\rm c} \geq \alpha \geq 0$. 
The curve $H^{\rm 1D}_{\rm s}(\alpha)$ is {\it not} smooth: it consists of quasi-linear parts 
with an infinite number of slope jumps at the endpoints of each quasi-linear part. 
These intervals become shorter and shorter when $\alpha \rightarrow \alpha_{\rm c}$ 
(see Fig.\ S3 of Ref.\ \onlinecite{epaps}). These non-analytic endpoints reflect 
the changes of multi-MBS related low-energy excitations. This specific behavior persists also 
for $D=2, 3$ (see Figs.\ 4, 5) and is a signature of quantum  phase transitions. 
Concerning \Litwo \ we stress that a 1D-approach yields for $\alpha=0.332$ \cite{Lorenz09}, 
$h_{\rm s}=0.0616916$ (see Table I) where $g=2$ and $J_1=228$~K have been used.
Thus, the empirical value $H_{\rm s}= 55.4$~T reported below is strongly {\it underestimated} 
by a 1D-approach. Hence, let us  consider, what happens, if an IC of the type shown in Fig.\ 1 
is switched on. In general, a complex phase diagram in terms of the IC 
and $\alpha$ has been obtained (see Fig.\ 3 and Ref.\ \onlinecite{epaps}). First, with increasing 
IC above a critical $J^{\rm cr,-}_{\rm ic}(\alpha)$ the MP phase is removed in favor 
of one of the two incommensurate (INC) dipolar phases where $H_{\rm s}$ (given exactly by the SWT) 
is still affected by both in-chain and IC although the contribution of the former for $\alpha<0.54$ 
is significantly reduced as compared to the 1D case (see Figs.\ 4, 5). But at stronger IC 
for $\alpha<1$ one reaches a second critical point $J^{\rm cr,+}_{\rm ic,2}$ where the 
INC phases are suppressed in favor of a commensurate (C) phase with FM in-chain
correlations (see Figs.\ 3-5). 
Then $H_{\rm s}$ depends {\it solely} on the IC:
\begin{equation}
g \mu_{\mbox{\tiny B}} H_{\rm s} =  N_{\rm ic} \left(J_{\rm ic}+J^\prime_{\rm ic}\right)
\ \mbox{for} \ J_{\rm ic} \geq J^{\rm cr,+}_{\rm ic},
\label{hsatli}
\end{equation}
where $N_{\rm ic}$ is the number of nearest interchain neighbors (8 in case of Li$_2$CuO$_2$). 
For $\alpha>1$ there is no C-phase. Now we take $J^\prime_{\rm ic}=0$ for the sake of simplicity.
Taking into account it 
%Its account 
is possible  but electronic structure calculations suggest $J^\prime_{\rm ic}/J_{\rm ic} \leq 0.1$, 
only. Then, $J^{\rm cr,+}_{\rm ic}(\alpha) = (4\alpha -1)|J_1|/9$, 
if $\alpha<0.57$ which is relevant for \Litwo. For the remaining interval $0.57 \leq \alpha<1$, 
the INC2-C transition is of 1$^{\rm st}$ order, see Ref.\ \onlinecite{epaps}.
Turning to \Litwo \ and ignoring a small 
%spin 
anisotropy \cite{Lorenz09}, we have 
$J_{\rm ic} \approx 9.04 \mbox{K} > J^{\rm cr,+}_{\rm ic}(0.332)=0.0364|J_1| \approx 8.2$~K \cite{remerror}. 
Since in the C-phase $H_{\rm s}$ depends solely on $J_{\rm ic}$, the
IC can be directly read off 
from its measured value $H_{\rm s}$=55.4~T yielding $J_{\rm ic}=9.25$~K very close to $9.04$~K from
zero-field INS data \cite{Lorenz09}. In the INC1 phase $J_{\rm ic}$ dominates $H_{\rm s}$. 
There
%In this phase 
above 
$J^{\rm cr,-}_{\rm ic}(0.332) \approx 0.0109 |J_1|=2.49$~K only INC 1-magnon low-energy excitations 
exist. Below $J^{\rm cr,-}_{\rm ic}$ 3-MBS are recovered as low-energy excitations.
The transition from the 3-MBS- to the INC1-phase is 1$^{\rm st}$ order. The general situation is 
shown in Fig.\ 4. Well below $J^{\rm cr,+}_{\rm ic}$ the saturation field significantly depends
on $\alpha$. The yellow stripe highlights
the region where the IC is predominant as
%of predominant IC 
addressed in our title. 
Here this dependence is weak and smooth. Subtracting their classical value, the spin-spin
correlation functions show that the in-chain fluctuations vanish much faster than the  IC 
ones for $H \to H_{\rm s}$ (see Fig.\ 4). This explains the surprising result of Eq.\ (\ref{hsatli}), 
too. For systems in the FM subcritical in-chain frustration regime $\alpha<1/4$ the in-chain contribution 
vanishes by definition. Hence, the external field has to overcome the AFM IC, only, and 
can be used to extract the IC exactly, if the corresponding $g$-factor is known (see Eq.\ (1)). 
This case is realized  e.g.\ for \Catwo \ \cite{Kudo05}. It has a similar IC geometry 
as \Litwo, but a 2D arrangement of chains, i.e.\ $N_{\rm ic}=4$. This reduced $N_{\rm ic}$ is 
overcompensated by a larger IC $J_{\rm ic}+J'_{\rm ic}=26$~K \cite{Matsuda2001}. 
As a result we predict $H_{\rm s} \approx 61$~T and refine a recent overestimated value
of $70$~T  from extrapolated low-field data \cite{Kudo05}.

\begin{figure}[t]
\includegraphics[width=7.0cm]{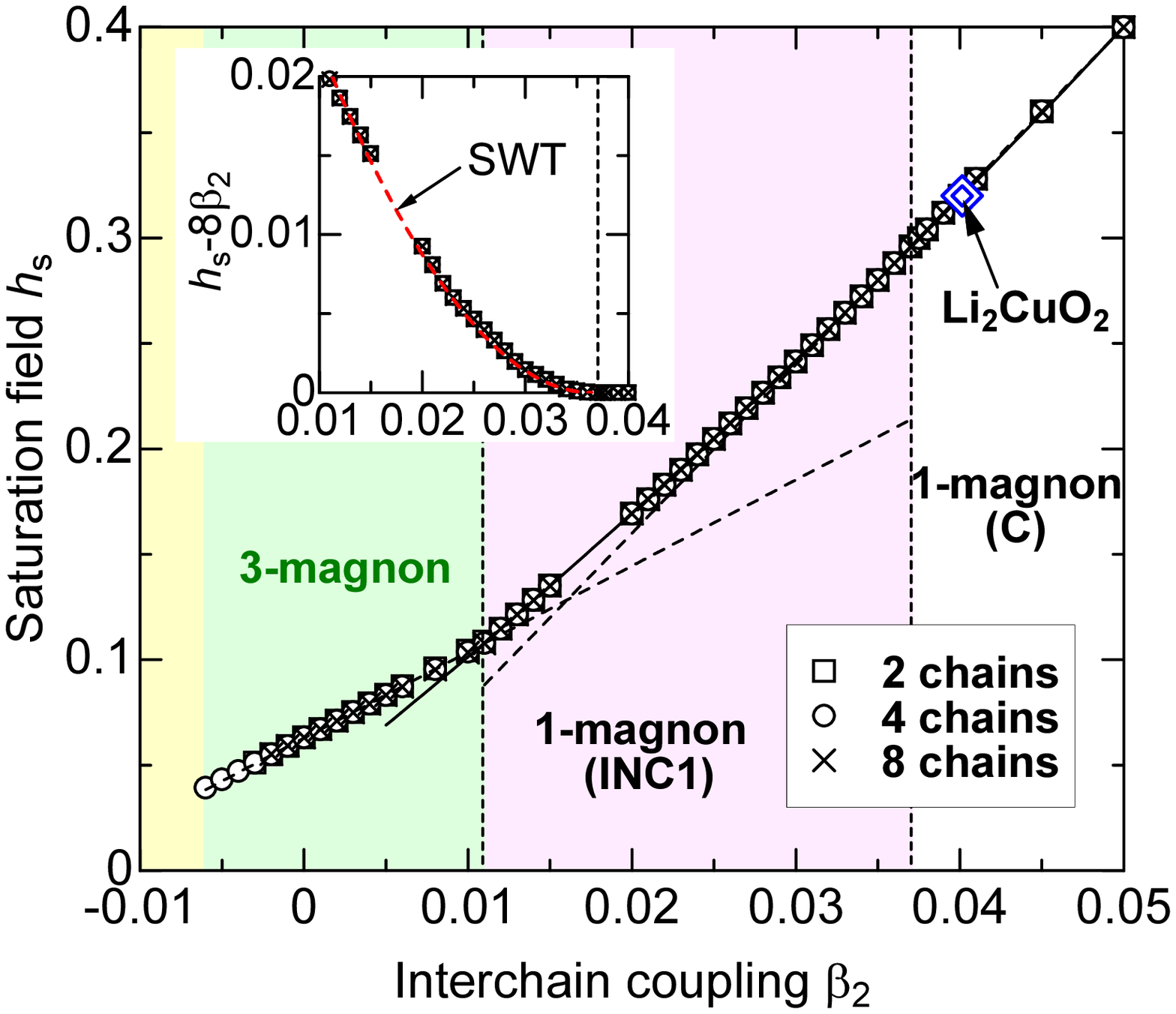}
\caption{(Color) 
The saturation field $h_{\rm s}$ at $\alpha=1/3$ vs.\ $\beta_2$ from DMRG method. 
Notice the two critical IC values $\beta^{\rm cr,-}_2=J^{\rm cr,-}_{\rm ic}/|J_1|=0.0109$ and 
$\beta^{\rm cr,+}_2=J^{\rm cr,+}_{\rm ic}/|J_1|=0.03704$ denoted by dashed vertical lines. 
For INC1- and C-phases, see text. Inset: the weak non-linearity of $h_{\rm s}$ vs.\ $\beta_2$.
 } 
\label{f5}
\end{figure}

\begin{figure}[t]
\includegraphics[width=8.5cm]{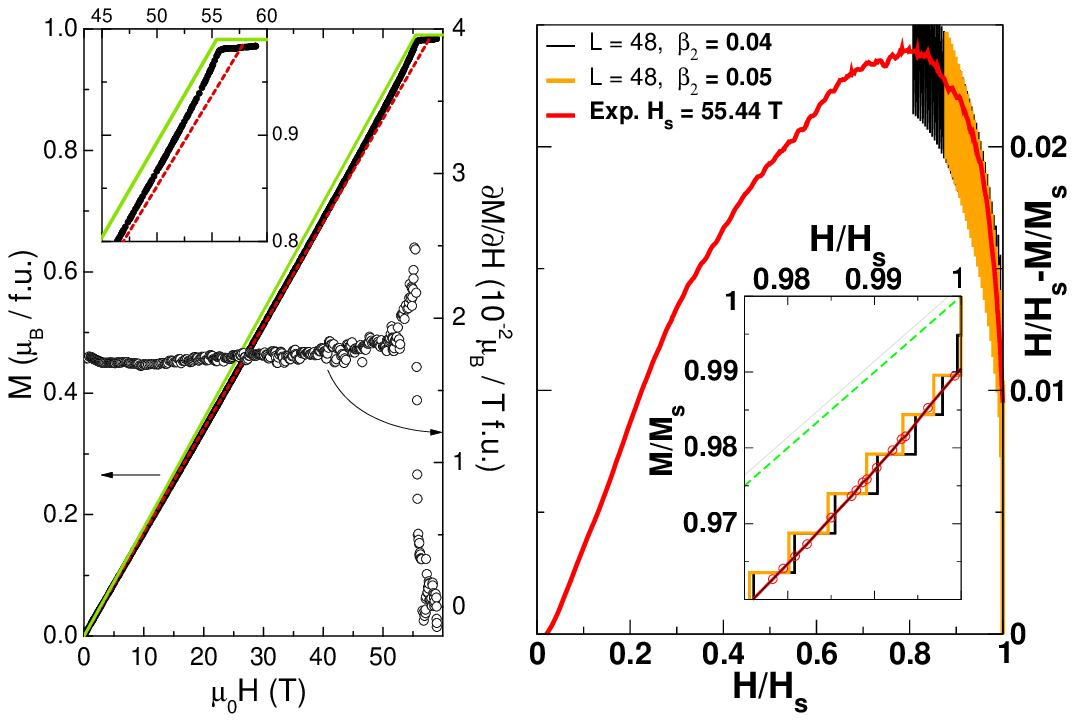}
\caption{(Color) Left: Magnetization $M$ (left axis) and the derivative $\partial M/\partial H$ ($\circ$) 
from pulsed fields. Red (green dashed) line: auxiliary line to make the upturn near $H_{\rm s}$ 
better visible (classical curve (CC): $M/M_{\rm s}=H/H_{\rm s}$).
Right: Deviations from CC: DMRG vs.\ experiment for different IC $\beta_2$ and a $H_{\rm s}$ value 
within the experimental error bars. Inset: Behavior just below $H_{\rm s}$ for the two $\beta_2$-values 
shown in the main part.
}
\label{puls}
\end{figure}

Pulsed-field magnetization studies 
%measurements 
have been performed at the Dresden High Magnetic Field Laboratory 
in fields up to $60$~T. The results taken at $T=1.45$~K for $H \parallel$ {\it b}-axis on a \Litwo \ single 
crystal from the same batch as in the INS-study \cite{Lorenz09} is shown in Fig.~\ref{puls}. 
The data imply a quasi-linear increase of the magnetization $M(H)$ between $10$ and $30$~T, i.e.\ 
$\partial M/\partial H=M^\prime \approx$ const. Above about $50$~T $M'$ increases notably and  
pronounced peaks develop at $55.4 \pm 0.25$~T and $55.1 \pm 0.25$~T for two pieces of our single crystal. 
The sharp drop of $M^\prime$ towards $0$ at higher fields justifies to attribute the peaks with 
$H_{\rm s}$. The saturation moment amounts to $M_{\rm s}\approx 0.99 \pm 0.06 \mu_{\mbox{\tiny B}}$/f.u. 
using $g_{\mbox{\tiny \it b}}=1.98 \pm 0.12$ in reasonable agreement with $g_{\mbox{\tiny \it b}}=2.047$ 
from low-field ESR-data at $300$~K \cite{Kawamata04}. 
In Fig.\ \ref{puls} we compare 
$M/M_{\rm s}$
%the reduced magnetization 
from the DMRG with the experiment for \Litwo. 
The DMRG description in this standard plot is rather good, but it is inconvenient to extract  
parameter values. This is due to the  weak  quantum fluctuations in  \Litwo. The function
$f(M)=H/H_{\rm s}-M/M_{\rm s}$ reflects the remaining quantum fluctuations much better (see Fig.\ 6). 
Notice the enhanced fluctuations (i.e.\ smaller $M(H)$) for the weaker IC (see inset). 
%We remind the reader that 
Noteworthy,
the 'width' of the DMRG curves 
%in Fig.\ 6 comes 
stems
from the finite 
%magnetization 
steps of $M(H)$ 
due to
%caused by 
the finite $L$. 
The almost straight shape of the $M(H)$ curves shown in Figs.\ 2 and 6 evidences the 3D character 
of \Litwo \ in accordance with the large local magnetic moment $\approx 0.93 \mu_{\mbox{\tiny B}}$ observed in the ordered phase
at low $T$ at $h=0$. The measured $H_{\rm s}$ of our single crystal results in $J_{\rm ic}=9.25 \pm 0.04$~K, 
%very 
close to the INS-data. Finally, we note that linear 
%combinations 
relations
of two experimentally
accessible thermodynamic quantities, the Curie-Weiss temperature and $H_{\rm s}$, yield
useful constraints for the in-chain $J$'s 
%(see Ref.\ 
\cite{epaps}.

%}) .

To summarize, the crucial role of  AFM IC in frustrated quasi-1D systems, such as 
\Litwo, for their behavior in external  fields and, particularly, the strength of $H_{\rm s}$ 
has been demonstrated. Extracting $J_{\rm ic}$ for that 
%compound 
system
from pulsed-field data,  
we arrived at $9.25$~K, very close to a previous INS study which confirms the validity 
of the adopted spin model. To extract $J_{\rm ic}$ a $H_{\rm s}$-study is preferable 
over an INS, due its smaller error bars, the possibility to work with small single crystals 
and its much higher efficiency (concerning both time and costs). The large $H_{\rm s}$ from 
%two
completely different studies discards any 1D scenario for \Litwo, even more, $H_{\mbox{\tiny s}}$ 
itself is (within the isotropic model) independent of the in-chain couplings $J_1$ and $J_2$. 
Thus, our results show {\it exactly} that for a relatively wide interval $0<\alpha<1$ and 
collinear magnetic 
%long-range 
order at $H=$0, $H_{\rm s}$ depends only on the IC, 
irrespectively of its strength. A complete study of the 
entire 
phase diagram including the 
INC2-phase 
which at present is rather
of academic interest, only, will be 
%given
presented elsewhere. 
The MP-phases from 1D studies are very sensitive to the presence of IC in the 2D/3D counter parts. 
In particular, they can be readily eliminated by a weak AFM IC. Instead 
%two 
new incommensurate 
phases may occur. A study of other 
%compounds 
systems
and their $H_{\rm s}$ within the approach 
proposed here is promising and in progress.

We thank the DFG (grants DR269/3-1, KL1824/2, RI615/16-1),
PICS program (contr.\ CNRS 4767, NASU 267), EuroMagNET, \& 
EU (contr.\ 228043) for support.\\

\vspace{-0.1cm}

\noindent
$^*$ Corresponding author: s.l.drechsler@ifw-dresden.de

\newpage

\begin{widetext} 

\large
\begin{center}
{\bf EPAPS supplementary online material:\\
\char`\"{}Saturation field of frustrated chain cuprates: regions of predominant
interchain coupling\char`\"{}}
\end{center}

\normalsize

\vspace{1.0mm}
\begin{center}
S.~Nishimoto$^1$, S.-L.~Drechsler$^1$, R.O.\ Kuzian$^1$, J.\ van den Brink$^1$\\
J.~Richter$^2$, W.E.A.~Lorenz$^1$, Y.\ Skourski$^3$, R.~Klingeler$^4$, B.~B\"uchner$^1$\\
{\it
$^1$IFW Dresden, P.O.~Box 270116, D-01171 Dresden, Germany\\
$^2$Universit\"at Magdeburg, Institut f\"ur Theoretische Physik, Germany\\
$^3$Hochfeld-Magnetlabor Dresden, Helmholtz-Zentrum Dresden-Rossendorf, D-01314 Dresden, Germany\\
$^4$Kirchhoff Institute for Physics, University of Heidelberg, D-69120 Heidelberg, Germany\\
}
\end{center}

\renewcommand{\theequation}{S\arabic{equation}}

\renewcommand{\thefigure}{S\arabic{figure}}

\renewcommand{\thetable}{S\arabic{table}}

\setcounter{equation}{0}

\setcounter{figure}{0}

\setcounter{table}{0}

%\newpage

\vspace{1.5cm}

In the present Supplementary part we provide the reader with the general
phase diagram of the 3D model with a diagonal 
antiferromagnetic (AFM) interchain coupling (IC) $J_{\rm ic}$ that couples 
each site in a frustrated $J_1$-$J_2$ single 
chain to its next-nearest-neighbor (NNN) sites 
in the adjacent chains.
This type of IC is realized in \Litwo, see Fig.~1 of the main text.
Its rigorous treatment leads for the incommensurate parts INC1
and INC2 of the general phase diagram (see Fig.\ S1)
to somewhat tedious analytical expressions for the saturation field $H_{\rm s}$
provided here for an interested reader explicitly.
Then, in the spirit of our proposed efficient thermodynamic approach, 
we present simple but exact equations where $H_{\rm s}$
together with the Curie-Weiss temperature can be used to get insight into 
the microscopic exchange couplings.
Finally, we present DMRG data mentioned in main text for the most realistic
case of a very weak nearest-neighbor (NN) 
IC $\beta_1=J'_{\rm ic}/|J_1|$
and a predominant NNN IC
$\beta_2=J_{\rm ic}/|J_1|$
with parameters relevant for \Litwo \ suggested by band structure calculations \cite{nitzsche11}.

\vspace{0.4cm}

\centerline{\bf PHASE DIAGRAMS AND THE DETERMINATION OF THE SATURATION FIELD}

\vspace{0.3cm} 

We start with the presentation of
the  whole phase diagram \cite{remark1}, see Fig.\ S1.

\begin{figure}[ht]
\includegraphics[width=7.5cm]{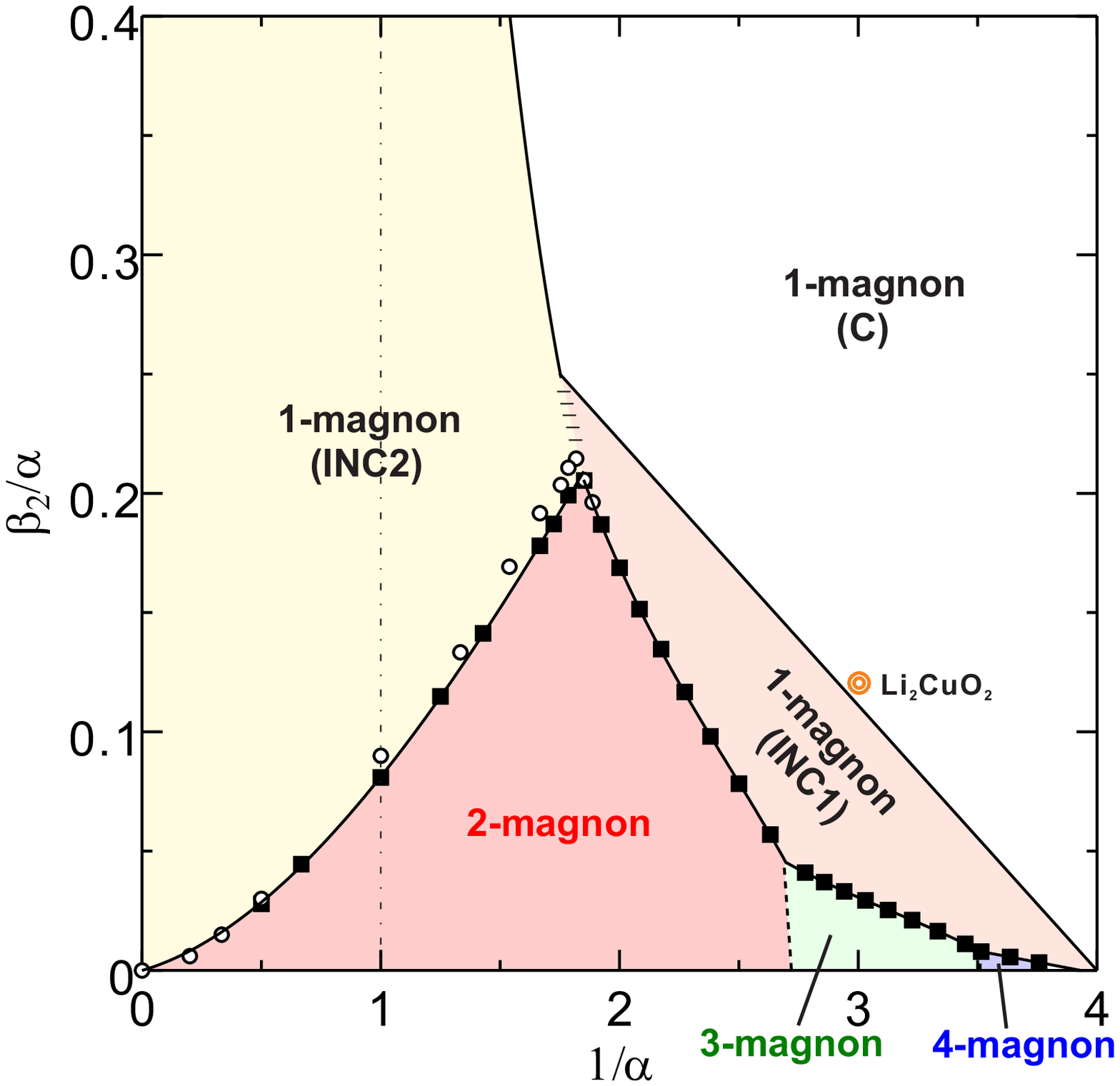}
\hspace{1.7cm}
\includegraphics[width=7.8cm]{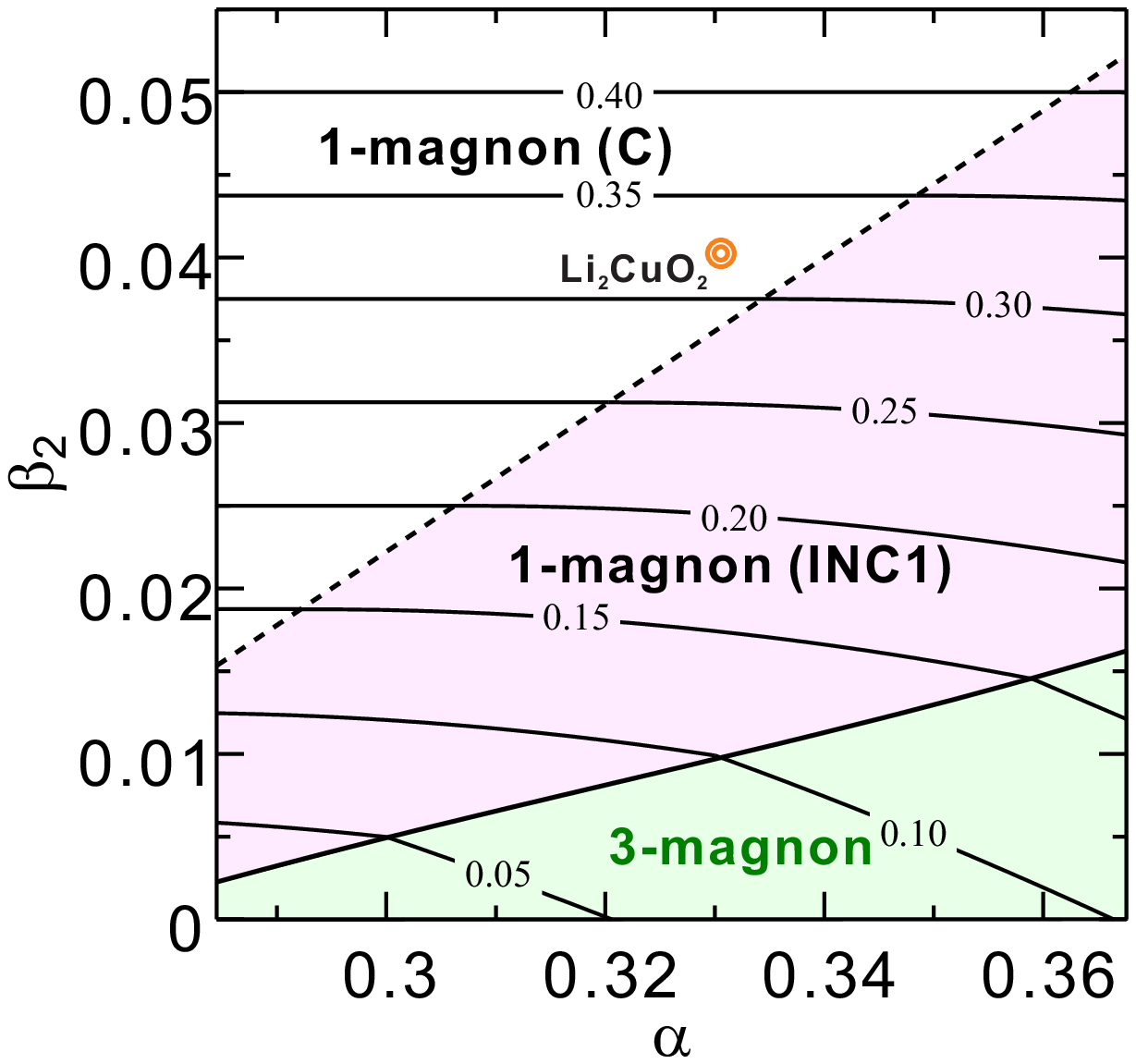}
\caption{(Color) Left: General phase diagram for frustrated chains with an AFM IC $\beta_2$ 
of the diagonal type as realized in \Litwo \ (see Figs.\ 1 and 3 of the main text) but
for arbitrary reciprocal frustration parameter $1/\alpha=|J_1|/J_2$. The symbol 
$\blacksquare$ denotes our DMRG results and the symbol $\circ$ stands for the exact
hard-core boson approach. Notice a second incommensurate phase INC2. Its border line 
to the commensurate (C) phase approaches asymptotically the value of $1/\alpha=1$ as shown 
by the dashed vertical line. For the behavior along the basis line, i.e. the 1D limit, 
see Fig.\ S2 and Refs.\ 15 and 16 in the main text. The line separating the two incommensurate 
one-magnon phases INC1 and INC2 represents a first-order transition.
Right: Enlarged contour plot of the dimensionless saturation field 
$h_{\rm s}=g\mu_{\mbox{\tiny B}}H_{\rm s}/|J_1|$ vs.\ the in-chain frustration rate 
$\alpha$ and the interchain interaction $\beta_2$ in the vicinity of 
parameters for \Litwo. The boundary between the INC1 and the 3-magnon phases is 
obtained by our DMRG calculations. Notice the slightly non-equidistant height differences 
in the contour lines of constant $h_s$ in the INC1 region as compared to equidistant ones 
in the C-region as well as the changed axis notation as compared with the left part. 
}
\label{f2}
\end{figure}

\begin{figure}[t]
\includegraphics[width=9cm]{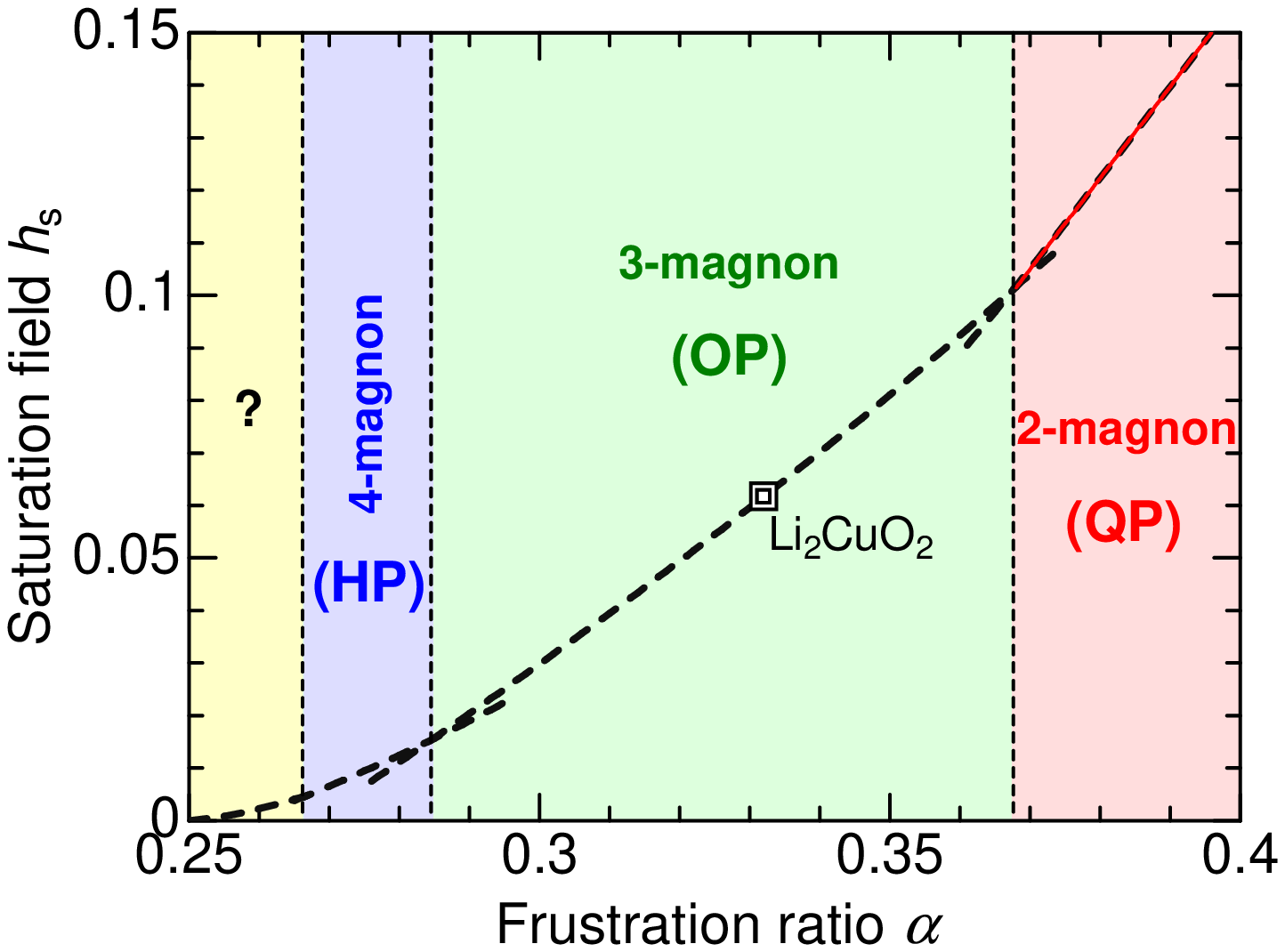} 
\caption{(Color) Saturation field $h_{\rm s}$ vs.\ frustration ratio
$\alpha$ of a single chain according to our DMRG. Notice the 
change in the slope indicated by the vertical dashed lines.
The octupolar (OP or 3-MBS) is realized for $0.2845\leq\alpha\leq0.3676$ 
in between
the hexadecupolar (HP or 4-MBS) and the quadrupolar (QP or 2-MBS) regions. 
Dashed line: DMRG results
(each dash contains about 5 data points). Red solid line: Analytical curve
from the exact 2-magnon solution obtained from the hard-core boson
representation of the general spin Hamiltonian (valid exactly
in the one- and two-magnon regions) given in Eq.\ (S1).
The question mark in the most left yellow region close to the critical
point $\alpha = 1/4$ stands for numerically still unresolved but 
expected higher order multipolar phases (see text).
}
\label{f2}  
\end{figure}

Let us first remind the behavior of the saturation field at the basis line of the general 
3D phase  diagram, i.e.\ the 1D limit of the considered problem, see Fig.\ S2
and also Refs.~15 and 16 in the main text. At present the higher multipolar phases beyond 
the four-magnon sector near the critical point $\alpha_{\rm c}=1/4$ cannot been assigned 
numerically with sufficient precision as a sequence of 5-, 6-, etc.\ phases as suggested 
by Krivnov and Dmitriev on general grounds \cite{DK06}. The question mark in Fig.\ S2 
reminds the reader of this numerically still somewhat unclear situation.

The hard-core boson approach provides an exact solution (see e.g.\ Ref.\ 11 in the main text) 
for the 2-magnon (nematic, quadrupolar (QP)) phase.  
In the 1D case we obtain in the QP phase
\begin{equation}
h_{\rm s}=2\alpha -1 +0.5/(\alpha+1),
\end{equation}
valid above a {\it critical } point $ \alpha_{\rm 3} \approx 0.3676776 $, i.e.\ for 
$\alpha \geq \alpha_{\rm 3}$. Below $\alpha_{\rm 3}$ at first 3-MBS and then 4-MBS, etc.\ occur
as the lowest excitations. In the 1D case our DMRG data for the 3-magnon (octupolar) phase 
can be described with high precision by an expansion up to the second order around the critical 
point $\alpha_{\rm 3}$
\begin{equation}
h^{\rm 3M}_{\rm s}(\alpha)=h_{\rm 03}+
h'_{\rm 3}(\alpha-\alpha_{\rm 3})+0.5 h''_3(\alpha-\alpha_{\rm 3})^2 + ...,
\label{a1}
\end{equation}
where $h_{\rm 03}=0.1009384$, $h'_{\rm 3}=1.1511$, and $h''_{\rm 3}=2.95699$. 
Eq.\ (S2) is valid for $\alpha_{\rm 3} \geq \alpha \geq \alpha_{\rm 4} \approx 0.284533$. 
A similar simple but semi-analytic solution, only, can be derived also for the 4-magnon 
phases based on a fit of our DMRG results.

The inspection of Fig.\ S3 shows that in the limiting case $\alpha=1$ at sizable IC, 
say for $J_{\rm ic}/J_{\rm 2} \geq 1/3$, the saturation field $H_s$ is also dominated 
by the IC. Note that a similar effect for a sizable IC appears also for $\alpha > 1$. 
But it becomes weaker with increasing $\alpha$.

\vspace{0.6cm}

\centerline{\bf THE 3D SPIN-HAMILTONIAN}

\vspace{0.3cm}
The spin-1/2 Heisenberg Hamiltonian reads
\begin{equation}
\hat{H}_{\rm H}=\sum_{\mathbf{R}}\left[J_1
\left(\mathbf{\hat{S}}_{\mathbf{R}}\cdot
\mathbf{\hat{S}}_{\mathbf{R+b}}\right)+
J_2\left(\mathbf{\hat{S}}_{\mathbf{R}}\cdot
\mathbf{\hat{S}}_{\mathbf{R+2b}}\right)+
J_{\rm ic}\sum_{\mathbf{r}}\left(\mathbf{\hat{S}}_{\mathbf{R}}\cdot
\mathbf{\hat{S}}_{\mathbf{R+r}}\right)+
J_{\rm ic}^{\prime}\sum_{\mathbf{r}^{\prime}}\left(\mathbf{\hat{S}}_{\mathbf{R}}\cdot
\mathbf{\hat{S}}_{\mathbf{R+r}^{\prime}}\right)
-\mu\left(\mathbf{H}\cdot\mathbf{\hat{S}}_{\mathbf{R}}\right)
\right],
\label{eq:Hs}
\end{equation}
where $\mathbf{R}$ is the position vector of a Cu site in the lattice,
$\mathbf{r=\left(3b \pm a\pm c\right)}/2$, $\mathbf{r^\prime=\left(b \pm a \pm c\right)}/2$
$\mathbf{a,b,c}$ are lattice vectors.

The one- and two-magnon excitation spectrum for magnetic fields $H>H_{\rm s}$ can be found 
{\it exactly} at $T=0$. The classical fully spin-polarized ground state is also an exact eigenstate 
of the quantum Hamiltonian. The spin-deviation (magnon) operators for the spin-1/2 Heisenberg 
Hamiltonian are Pauli (hard-core boson) operators, in terms of which the Hamiltonian is 
rewritten as \cite{Tiabl}
\begin{eqnarray}
\hat{H} & = & \hat{H}_{0}+\hat{H}_{\rm int},\label{Hviab}\\
\hat{H}_0 & = & \sum_{\mathbf{R}}\left[\omega_{0}\hat{n}_{\mathbf{R}}+
J_1b_{\mathbf{R}}^{\dagger}b_{\mathbf{R+b}}+J_2b_{\mathbf{R}}^{\dagger}b_{\mathbf{R+2b}}+
J_{\rm ic}\sum_{\mathbf{r}}b_{\mathbf{R}}^{\dagger}b_{\mathbf{R+r}}+
J_{\rm ic}^{\prime}\sum_{\mathbf{r}^{\prime}}b_{\mathbf{R}}^{\dagger}
b_{\mathbf{R+r}^{\prime}}\right]\:,\label{H0}\\
\omega_{0} & \equiv & \mu H-\left[J_1+J_2+4\left(J_{\rm ic}+
J_{\rm ic}^\prime\right)\right],\:\hat{n}_{\mathbf{R}}\equiv 
b_{\mathbf{R}}^{\dagger}b_{\mathbf{R}}\nonumber \\
\hat{H}_{\rm int} & = & \sum_{\mathbf{R}}\left[
J_1\hat{n}_{\mathbf{R}}\hat{n}_{\mathbf{R+b}}+
J_2\hat{n}_{\mathbf{R}}\hat{n}_{\mathbf{R+2b}}+
J_{\rm ic}\sum_{\mathbf{r}}\hat{n}_{\mathbf{R}}\hat{n}_{\mathbf{R+r}}+
J_{\rm ic}^{\prime}\sum_{\mathbf{r}^{\prime}}\hat{n}_{\mathbf{R}}
\hat{n}_{\mathbf{R+r}^{\prime}}
%-\mu\left(\mathbf{H}\cdot\mathbf{\hat{S}}_{\mathbf{R}}\right)
\right]\: ,
\label{Hint}
\end{eqnarray}
where $\mu =g\mu_{\mbox{\tiny B}}$ and the $z$-axis is parallel to $\mathbf{H}$. 
The transverse part of the Heisenberg Hamiltonian $\hat{H}$ (\ref{eq:Hs}) defines the one-particle
hoppings in $\hat{H}_0$ (\ref{H0}), the Ising part contributes the interaction (\ref{Hint}) and 
the on-site energy value $\omega_0$. Note that a negative (ferromagnetic) exchange in (\ref{Hint}) 
acts as an attraction between the magnons. In 1D $J_{1}$-$J_{2}$-chains with $J_1<0$ and $J_2>0$ 
this leads always to the formation of multimagnon bound-states. In 2D and 3D lattices the bound-states 
formation is possible only if the attraction overcomes the kinetic energy. Besides on the strength of 
the AFM interchain interaction it depends also on the number of available interchain neighbors which in 
the case of diagonal IC is doubled as compared with the frequently studied but much simpler case 
of perpendicular IC of unshifted chains (see Figs.\ 1 of the main text and of Ref.\ 17 therein).
Therefore, the case of a 3D chain arrangement as in \Litwo \ is especially effective in deconfining 
the multimagnon bound states and suppressing a spiral type 1-magnon state as the phase INC1 and INC2 
considered above.

The Hamiltonian (\ref{Hviab}) conserves the number of magnons. So,
the excitation spectrum may be found separately for one-, two-, three-
etc.\ magnon sectors of the Hilbert space. The saturation field is
determined by the condition of the stability of the ground state.
It becomes unstable when the frequency of a certain excitation vanishes.

\vspace{0.5cm}

\centerline{\bf THE TWO INCOMMENSURATE ONE-MAGNON PHASES INC1 AND INC2}

\vspace{0.3cm}

The energy of one-particle (1-magnon) excitations is particularly simple. For the isotropic model 
and $J^\prime_{\rm ic}=0$ it reads 
\begin{equation}
\omega_{\mathbf{q},\sigma}=\mu_{\mbox{\tiny B}}H_{s}+J_{1}\left(\cos q_{b}-1\right)+
J_2\left(\cos2q_{b}-1\right)+4J_{\rm ic}\left(\sigma\cos\frac{q_{a}}{2}
\cos\frac{3q_{b}}{2}\cos\frac{q_{c}}{2}-1\right),
\label{eq:wq}\end{equation}
where we have retained the AFM Brillouin zone, $\sigma\equiv\pm1$ enumerates the two branches
of the spectrum. The extrema of the one-particle spectrum are given by the equation 
$\nabla\omega_{\mathbf{q}}=0$, which gives for our model $q_a=q_c=0$, and 
\begin{equation}
\sin
\frac{q_b}{2}
\left[
J_1
\cos
\frac{q_{b}}{2}+4J_2
\left(
2\cos^{3}\
\frac{q_b}{2}
-\cos
\frac{q_b}{2}
\right)
+3\sigma J_{\rm ic}
\left(
4\cos^{2}
\frac{q_b}{2}-1
\right)
\right] 
= 0.
\nonumber
\end{equation}
The extrema for the two branches occur at $q_{b,\sigma}=0$ and at incommensurate
values of $q_{b}=q_{i,\sigma},\, i=1,2,3$, where the expression in square brackets vanishes
\[
\cos\frac{q_{i,\sigma}}{2}=y_{i,\sigma}-\sigma\frac{J_{\rm ic}}{2J_2},
\]
where $y$ is the root of the cubic equation 
\begin{eqnarray}
y_{1,\sigma} & = & 2\sqrt{-\frac{p}{3}}\cos\frac{\gamma_{\sigma}}{3},\quad 
y_{2,3,\sigma}=-2\sqrt{-\frac{p}{3}}\cos\frac{\gamma_{\sigma}\pm\pi}{3}\\
\cos\gamma_{\sigma} & \equiv & -\frac{\sigma q}{2\sqrt{\left(-p/3\right)^{3}}},\\
p & \equiv & -\frac{1}{2}\left[1-\frac{J_{1}}{4J_{2}}+
\frac{3J_{\rm ic}^{2}}{2J_{2}^{2}}\right],\\
q & \equiv & -\frac{J_{\rm ic}}{2J_{2}}\left[p+
\frac{1}{4}\left(\frac{J_{\rm ic}}{J_{2}}\right)^{2}+\frac{3}{4}\right].
\end{eqnarray}
In the dipolar phase, the saturation field $H_{\rm s}$ is determined by the condition that 
the energy of the lowest one-magnon excitation $\omega_{\mathbf{q},\mathrm{min}}$ vanishes.
Then, from Eq.\ (\ref{eq:wq}) the dimensionless saturation field 
$h_{\rm s}=g\mu_{\mbox{\tiny B}}H_{\rm s}/|J_1|$ is determined by the equation 
\begin{equation}
h_{\rm s}=\left(\cos q_{b,\mathrm{min}}-1\right)-
\alpha\left(\cos2q_{b,\mathrm{min}}-1\right)-
4\beta\left(\sigma\cos\frac{3q_{b,\mathrm{min}}}{2}-1\right) .
\label{eq:hsq}
\end{equation}
For an isolated chain ($\beta_2 =0$) the two branches are degenerated
and both reach the minimum at $q_{0}(0)=\arccos(1/4\alpha)$. 
For finite $J_{\rm ic}$ and $\alpha\lesssim0.57$ (the INC1 phase)
the minimum occurs at $q_{1,-1}$ of the branch $\sigma=-1$. With
the increase of $\beta_2$ the minimum shifts towards the centrum of
BZ, which is reached at $\beta_2=\beta^{\rm cr}_2(\alpha)=(4\alpha-1)/9$.

Now we note that $0<\beta_2 \leq \beta^{\rm cr}_2 \leq 4/9$ holds for all frustration 
values $1/4\leq\alpha<0.57$. Then the expression (\ref{eq:hsq}) may be expanded in powers of 
$\beta^{\rm cr}_2-\beta_2$
\begin{eqnarray}
h_{\rm s}(\alpha,\beta_2) & = & 8\beta_2-
\frac{7776\left(1-20\alpha\right)\left(\beta^{\rm cr}_2-\beta_2
\right)^{2}}{a^{2}}\left[h_{0}+h_{1}\left(\beta^{\rm cr}_2-\beta_2\right)+
h_2\left(\beta^{\rm cr}_2-\beta_2\right)^{2}\right],\label{eq:hsa}\\
a & \equiv & 5-72\alpha+1744\alpha^{2},\nonumber \\
h_0 & \equiv & \frac{1}{2}\left(5-72\alpha+592\alpha^{2}\right),\\
h_1 & \equiv & \frac{18\left(7-84\alpha+1184\alpha^{2}\right)
\left(1-20\alpha\right)}{a},\\
h_2 & \equiv & \frac{324\left(5-45\alpha+1204\alpha^{2}\right)
\left(1-20\alpha\right)^{2}}{a^{2}} .
\end{eqnarray}
For $\beta_2>\beta^{\rm cr}_2$ the minimum persists at $q_{b,-1}=0$,
and the saturation field is simply given by $h_{\rm s}=N_{\rm ic}\beta_2=8\beta_2$. 
For \Litwo \ one has $\beta_2 \approx3/76\approx0.03947>\beta^{\rm cr}_2=0.0364$. 
For $0.57<\alpha<1$ the minimum occurs at $q_{1,+1}$ of the branch 
$\sigma=+1$ for $\beta$ smaller a certain value $\beta^{\rm cr}_2(\alpha)>(4\alpha-1)/9$
where it jumps at $q_{b,-1}=0$. So, for this range of the frustration ratio $\alpha$, 
the transition between the commensurate C-phase with FM in-chain ordering
and the new incommensurate INC2 phase is of  first order. The corresponding 
phase boundary may be found from the condition $\omega_{+1}(q_{1,+1})=\omega_{-1}(0)$, 
and it is given by the following parametric function
\begin{eqnarray}
\alpha(t) & = & \frac{1-t+\frac{2\left(t-2\right)\left(2t-1\right)}{2t-7}}{4t\left[t\left(t-1\right)+\frac{2\left(2t-1\right)^{2}}{2t-7}\right]},\\
\beta^{\mbox{\tiny cr}}_2(t) & = & \frac{4t\alpha(t)\left(2t-1\right)+t-2}{\left(2t-7\right)\left(2t-1\right)}.
\end{eqnarray}
Finally, for $\alpha\geq1$ the minimum resides at $q_{1,+1}$ for
all values of IC. The resulting phase diagram is shown schematically
in Fig.\ S1 (left).

\vspace{0.4cm}

\centerline{\bf EXACT THERMODYNAMIC CONSTRAINTS INVOLVING THE SATURATION FIELD}

\vspace{0.3cm}

To demonstrate the general applicability of our proposed efficient thermodynamic
method to extract exchange integrals from simple and "cheap" measurements, 
we present a collection of simple but rigorous constraints valid for
situations where to first approximation a chain-like compound can be
described by the adopted isotropic $J_1$-$J_2$-model for a single chain with
one main interchain coupling, only,
but with various geometries of the
chain arrangement.
Generalizations to several
interchain couplings and to models with spin anisotropy are 
straightforward and will be given elsewhere.

With the experimental Curie-Weiss temperature $\Theta_{\mbox{\tiny CW}}$ (
that governs the high-$T$ spin susceptibility $\chi(T)\sim1/(T-\Theta_{\mbox{\tiny CW}})$)
which is simply related to the Hamiltonian parameters by
\begin{equation} \label{CW}
\Theta_{\mbox{\tiny CW}}\approx-0.5\left[J_1+J_2+0.5N_{\rm ic}\left(J_{\rm ic}+J'_{\rm ic}\right)\right]
\end{equation}
the whole set of interchain couplings between NN chains, including in particular
$J_{\rm ic}$ as well as all (weak) additional terms such as $J'_{\rm ic}$ etc.\ 
can be replaced by $\Theta_{\mbox{\tiny CW}}$
and Eq.\ (1) of the main text. As a results the we find 
a useful  empirical constraint for the in-chain exchange couplings $J_1$ and $J_2$.  
\begin{equation}
g\mu_{\mbox{\tiny B}}H_{\rm s}+4\Theta_{\mbox{\tiny CW}}=2|J_{1}|\left(1-\alpha\right).
\label{cwh}
\end{equation}
Using the experimentally measured saturation field for \Litwo \ 
of about $55.4$~T as reported in the main text, 
we predict a ferromagnetic Curie-Weiss temperature of about $55$ K
to be derived from high temperature ($T > 400$~K) susceptibility 
measurements not yet performed. For a strictly perpendicular 
NN IC, only, (see e.g.\ Fig. 1 in Ref.\ 17 ) 
the strong influence of the interchain coupling on the in-chain ordering 
is much reduced and as a result the 3D(2D) C-phase is missing, i.e.\ 
there is only one critical IC separating the {\it single}
INC phase from the MP phases.
In particular, the two-magnon phase shows a round 
maximum \cite{nishimoto10,ueda10} in the 3D case near $1/\alpha=0.906$ 
instead of the cusp seen in figure S1 (left) near $1/\alpha=1.754$. 
In the neighboring  single INC-phase $h_{\rm s}$ reads 
\begin{equation}
h_{\rm s}=2\alpha-1+0.125/\alpha+N_{\rm ic}J_{\rm ic} ,
\end{equation}
where $N_{\rm ic}$=2 (4) for a 2D (3D) chain arrangement, respectively.
Then Eq.\ (\ref{cwh}) is replaced by 
\begin{equation}
g\mu_{\mbox{\tiny B}}H_{\rm s}+4\Theta_{\mbox{\tiny CW}}=
|J_1|\left[1+0.125/\alpha\right] \ .
\end{equation}
In the two magnon-phase for weak IC one has 
\begin{equation}
g\mu_{\mbox{\tiny B}}H_{\rm s}+2\Theta_{\mbox{\tiny CW}}=|J_{1}|\left[\alpha+0.5/(1+\alpha)\right]
\end{equation}
(note the new prefactor of 2 in front of $\Theta_{\mbox{\tiny CW}}$).
Similar simple constraints can be derived for any  type of interchain couplings.
They provide another part for the proposed thermodynamic approach to extract
microscopic exchange integrals from few experimental numbers, only, instead to
fit for instance 
approximate expressions for complex quantities such as the sussceptibility 
or the magmetic spefic heat in more or less extended wide temperature 
intervals.

\vspace{0.6cm}

%\newpage

\centerline{\bf SOME DETAILED DMRG AND HARD-CORE BOSON RESULTS }

\vspace{0.3cm}

Finally, we present numerical DMRG  data  to demonstrate the extremely 
fast convergency
with respect to the number of chains taken into account.
Notice that here also a weak NN interchain coupling $\beta_1$
has been taken into account. The DMRG result is very close to
the analytical result of 0.350877192982456 given by Eq.\ (1) in the main
text.

\vspace{0.2cm}

\begin{table}[h]
\caption{Saturation field $h_{\rm s}$ at $J_1=-1$, $\alpha=0.332$, $\beta_2=3/76$,
$\beta_1=1/228$. $\beta_2$ and $\beta_1$ are multiplied by
four and two for 2-chain and 4-chain clusters, respectively.}
\begin{centering}
\begin{tabular}{cccccccc}
\hline 
$L$  & single chain  & 2-chain cluster  & 4-chain cluster  & 8-chain cluster  &  &  & \tabularnewline
\hline 
16  & 0.0610480058942  & 0.350877192884  & 0.350877192884  & 0.350877192884  &  &  & \tabularnewline
20  & 0.0618650864275  & 0.350877192884  & 0.350877192884  & 0.350877192884  &  &  & \tabularnewline
24  & 0.0616475064797  & 0.350877192884  & 0.350877192884  & 0.350877192884  &  &  & \tabularnewline
28  & 0.0617013393558  & 0.350877192884  & 0.350877192884  & 0.350877192884  &  &  & \tabularnewline
32  & 0.0616885713591  & 0.350877192884  & 0.350877192884  & 0.350877192884  &  &  & \tabularnewline
36  & 0.0616916345343  & 0.350877192884  & 0.350877192884  & 0.350877192884  &  &  & \tabularnewline
40  & 0.0616909145148  & 0.350877192884  & 0.350877192884  & 0.350877192884  &  &  & \tabularnewline
44  & 0.0616910784629  & 0.350877192884  & 0.350877192884  & 0.350877192884  &  &  & \tabularnewline
48  & 0.0616910423378  & 0.350877192884  & 0.350877192884  & 0.350877192884  &  &  & \tabularnewline
96  & 0.0616910487270  & 0.350877192884  &  &  &  &  & \tabularnewline
144  & 0.0616910487247  & 0.350877192884  &  &  &  &  & \tabularnewline
\hline
\end{tabular}
\par\end{centering}
\end{table}

In order to illustrate another aspect of the accuracy of our DMRG calculations,
we briefly discuss the difficult gap problem at $H=0$ of a single frustrated chain with 
a FM $J_1$ raised by recent field theory approaches \cite{nersesyan98,itoi01,berg07}. 
These theories predict a gap $\Delta$ of unknown magnitude. Since at present there are 
no hints from numerical studies for such a behavior, it is generally accepted
that such a gap should be very small \cite{berg07}. According to our DMRG calculations
we estimate  10$^{-5}|J_1|$ as an upper bound. For the case of 
Li$_2$CuO$_2$ under consideration this corresponds to $0.01$ to $0.02$~K, only. 
Anyhow, such a tiny gap, if it at all exists, would be of academic interest, only.

Now we present some numerical results from exact relations
obtained in the hard-core boson approach.
In Fig.\ S3 \ several examples for the IC dependence of the saturation field $h_{\rm s}$ are presented. 
Similar dependencies are obtained for arbitrarily large in-chain frustration ratios $\alpha$. 
In the 3D case with a finite antiferromagnetic $J_{\rm ic}$ the saturation field $h_{\rm s}$ 
is always affected by the former in contrast to the in-chain coupling
which
%In the two incommensurate phases the
%in-chain contribution is positive, too, although more or less reduced. 
can  
drop out exactly what happens in the commensurate C-phase. 
%For larger IC the 
%
%reduced in-chain 
%contribution to $h_{\rm s}$ is clearly seen in Fig.\ S3 (see also Fig.\ 4 in the main text). 
\begin{figure}[ht]
 %[htb]
\includegraphics[clip,width=0.45\columnwidth]{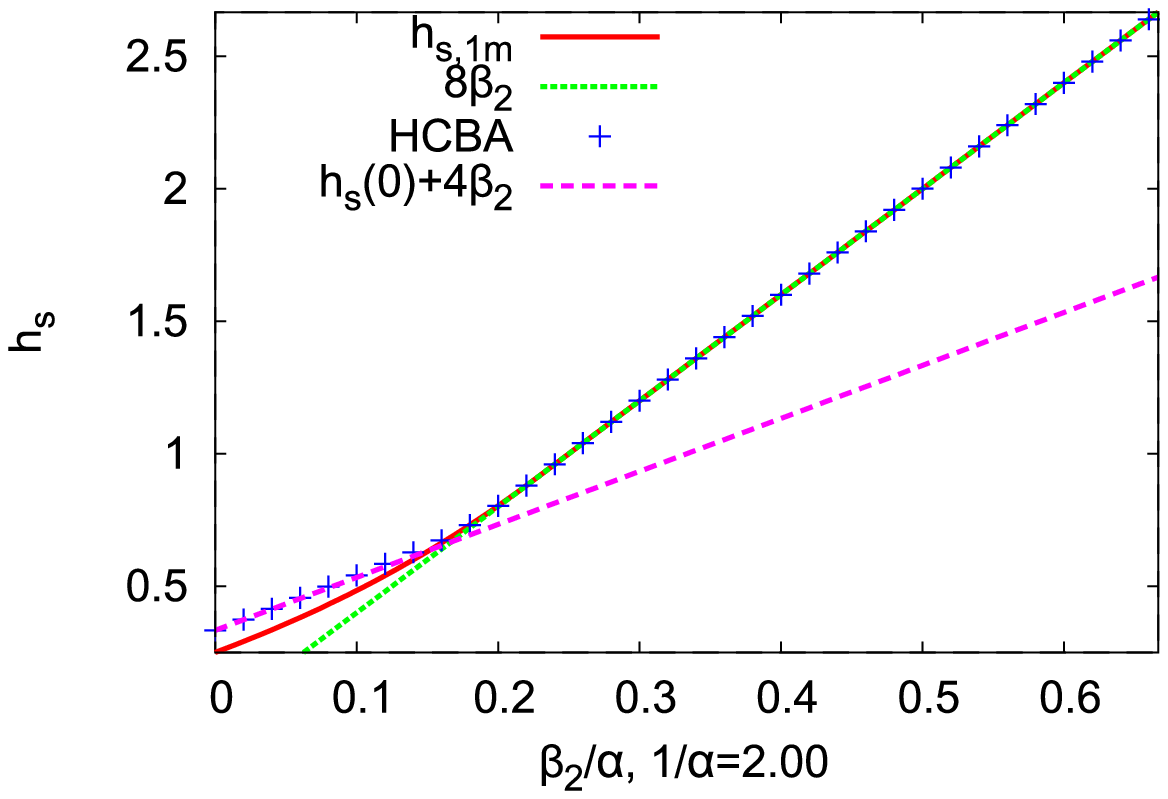}
\includegraphics[clip,width=0.45\columnwidth]{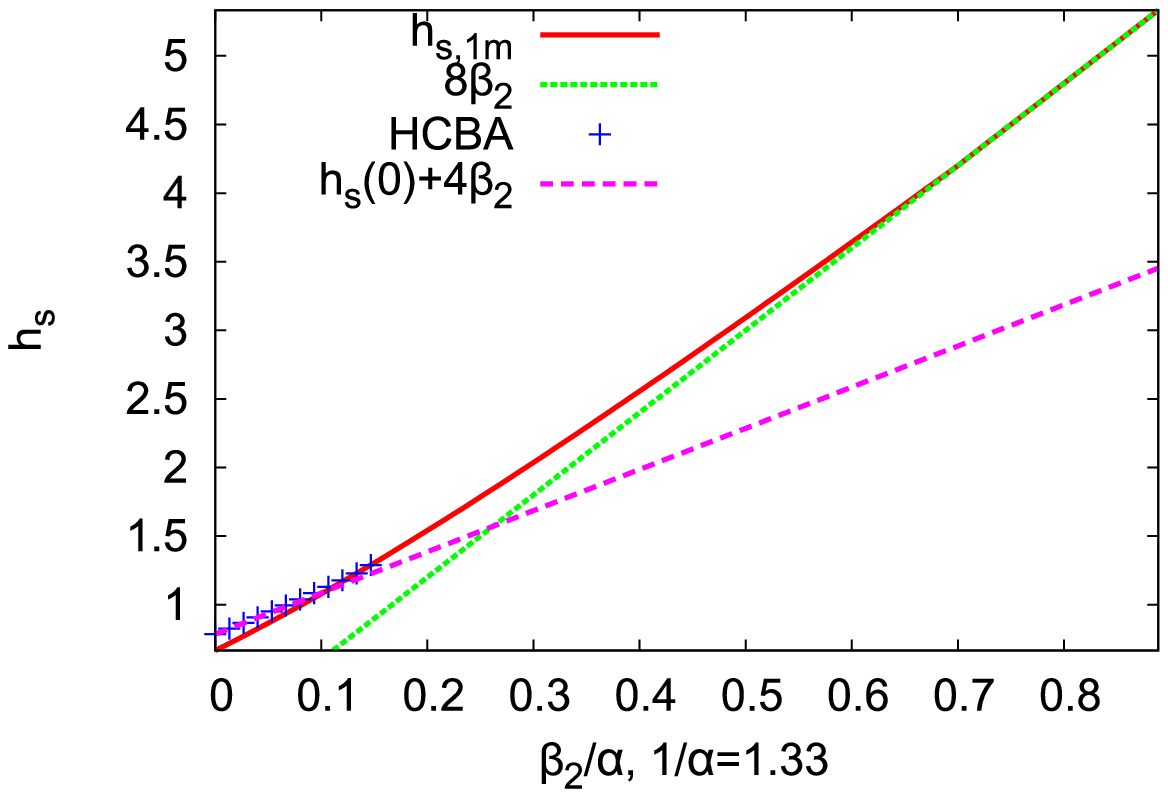}
\includegraphics[clip,width=0.45\columnwidth]{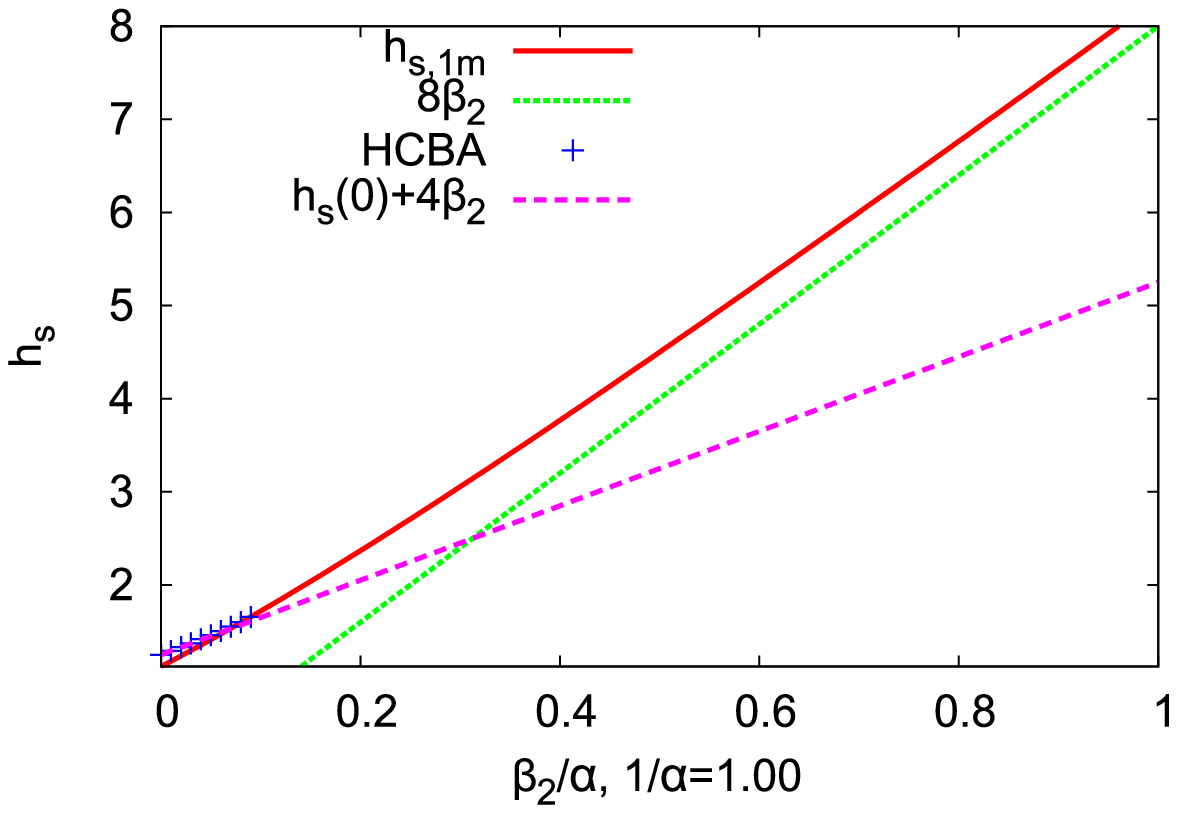} 
\caption{(Color) Saturation field for different in-chain frustration
rates $\alpha=0.5$, $0.75$, and $1.0$  vs.\ IC $\beta_2/\alpha=J_{\rm ic}/J_2$. 
The results of the exact hard-core boson approach are shown by symbols; the red solid line 
shows the one-magnon result, which becomes exact for 
$\beta_2 > \beta^{\rm cr,-}_2=J^{\rm cr,-}_{\rm ic}/|J_1|$ 
as explained in the main text (see also Figs.\ 3-5, therein). The green dashed line shows 
the value, which is reached in the commensurate C-phase, where the dependence on $J_1$, $J_2$ 
vanishes identically; the magenta thin line shows an approximation 
to the  exact solution in the 2-magnon phase. 
%Schematic high-field phase diagram of 
%the model (\ref{Hviab}), the labels C, INC1, and INC2 denote the region of the usual
%one-magnon dipolar phase, the "OP"-quadrupolar and the "OP"-octupolar phases, 
%respectively.
}
\end{figure}

\vspace{2cm}
\end{widetext} 


\begin{thebibliography}{99}
\bibitem{remerror} Considering the error bars of $J_1$ and $\alpha$: $\pm 5$~K and $\pm 0.005$, 
respectively, we see that $J^{\rm cr,+}_{\rm ic}$ is bounded by $7.18$~K 
$\leq J^{\rm cr,+}_{\rm ic} \leq 8.54$~K, i.e.\ $J^{\rm cr,+}_{\rm ic}$ is certainly below
$J_{\rm ic}=9.04 \pm 0.1$~K (INS) or $9.25 \pm 0.04$~K from our $H_{\rm s}$-data.
\end{thebibliography}
\begin{thebibliography}{99}
\bibitem{remark1}The actual numerical 
reasons for the rather small deviations between the DMRG results and 
those of the hard-core boson approach will be discussed in more detail 
elsewhere.
\end{thebibliography}
\end{document}